\documentclass[12]{article}         
\usepackage{amsfonts}
\usepackage{amssymb}
\usepackage{amscd}

\begin{document}
\title{Invariance in adelic quantum mechanics}  
%
\author{Branko Dragovich \\ Institute of Physics \\ Pregrevica 118, P.O.Box 57,  11001 Belgrade, Serbia}
%
%

\date{}
\maketitle

\begin{abstract}
Adelic quantum mechanics is form invariant under an interchange of
real and $p$-adic number fields as well as rings of $p$-adic
integers. We also show that in adelic quantum mechanics Feynman's
path integrals for quadratic actions with rational coefficients  are
invariant under changes of their entries within nonzero rational
numbers.
\end{abstract}

\section{Introduction}

Invariance of basic theoretical objects under some transformations
has played very important role in developments of quantum theory.
Quantum models  invariance is usually related to transformations in
Archimedean spaces characterized  by real or complex numbers.
However, since 1987 many models have been constructed over $p$-adic
and adelic spaces. In particular, much attention has been paid to
$p$-adic and adelic strings (see reviews in \cite{vladimirov1} and
\cite{freund}), $p$-adic \cite{vladimirov2} and adelic
\cite{dragovich1,dragovich2} quantum mechanics (see a review in
\cite{dragovich3}), $p$-adic and adelic quantum cosmology
\cite{dragovich4}, and to some systems which spaces of states
exhibit ultrametric hierarchical structures (see reviews
\cite{virasoro}, \cite{khrennikov} and \cite{kozyrev}). The present
status of application of $p$-adic numbers in physics and related
branches of sciences is reflected in the recent proceedings
\cite{proceedings}.

There are many mathematical and physical motivations
\cite{dragovich3} to employ $p$-adic numbers and non-Archimedean
geometry in modern mathematical physics. $p$-Adic and conventional
models are connected by the corresponding adelic models. Adelic
quantum mechanics (AQM) contains ordinary and $p$-adic quantum
mechanics in a natural way. It presents mathematically more complete
and theoretically more profound approach to quantum phenomena. Many
aspects of AQM have been investigated in detail \cite{dragovich3},
but a systematic analysis of symmetries was not considered.

 This paper contains a brief presentation of results obtained in a recent analysis
 of some invariance of AQM.

\section{$p$-Adic numbers and adeles}

$p$-Adic numbers are discovered at the end of the 19th century by
German mathematician Kurt Hensel. They can be obtained by completion
of the field $\mathbb{Q}$ of rational numbers in the same way as the
field $\mathbb{R}$ of real numbers, but using $p$-adic absolute
value instead of the ordinary one. $p$-Adic absolute value satisfies
strong triangle inequality $|x + y|_p \leq max \{ |x|_p \,, |y|_p
\}$ and consequently it belongs to non-Archimedean  (ultrametric)
norm. According to the Ostrowski theorem, $\mathbb{R}$ and all
fields $\mathbb{Q}_p$ of $p$-adic numbers, where $p$ is a prime
number, exhaust all number fields which can be obtained by
completions of $\mathbb{Q}$. In other words, only $\mathbb{R}$ and
$\mathbb{Q}_p$, for every $p$, contain $\mathbb{Q}$ as a dense
subfield.

Recall that a real number can be presented in the form
\begin{equation}
x = \pm 10^n  \, \sum_{k=0}^{-\infty} a_k \, 10^k\,, \quad a_k \in
\{ 0\,, 1\,, \cdots , 9 \} \,,  \quad  a_0 \neq 0\,,\,\,\, n\in
\mathbb{Z}\,, \label{2.1}
\end{equation}
where $\mathbb{Z}$ is the ring of rational integers. A $p$-adic
number has a unique expansion
\begin{equation}
y =  p^m  \, \sum_{k=0}^{+\infty} b_k \, p^k\,, \quad b_k \in \{
0\,, 1\,, \cdots , p-1 \} \,,  \quad  b_0 \neq 0\,,\,\,\, m\in
\mathbb{Z}\,. \label{2.2}
\end{equation}
It is evident that (\ref{2.2}) is not convergent with respect to the
usual absolute value, but it becomes quite  convergent applying
$p$-adic absolute value.

Due to ultrametricity, $p$-adic spaces have many properties rather
different from the real ones. For an introductory course to $p$-adic
numbers and $p$-adic analysis, one can see \cite{schikhof},
\cite{vladimirov1} and \cite{freund}.

 To consider real and $p$-adic numbers simultaneously and on an equal
footing one uses space ${\mathbb A}$ of adeles. An adele $x$ (see,
e.g. \cite{gelfand} and \cite{freund}) is an infinite sequence
\begin{equation}
  x= (x_\infty\,, x_2\,, \cdots, x_p \,, \cdots)\,, \label{2.3}
  \end{equation}
where $x_\infty \in {\mathbb R}$ and $x_p \in {\mathbb Q}_p$ with
the restriction that for all but a finite set $\mathcal P$ of primes
$p$ one has  $x_p \in {\mathbb Z}_p = \{y \in \mathbb{Q}_p\, : |y|_p
\leq 1 \} $. Componentwise addition and multiplication endow the
ring structure to  ${\mathbb A}$ and it can be presented in the
following form:
\begin{equation}
 {\mathbb A} = \bigcup_{{\mathcal P}} {\mathbb A} ({\mathcal P}),
 \ \ \ \  {\mathbb A} ({\mathcal P}) = {\mathbb R}\times \prod_{p\in
 {\mathcal P}} {\mathbb Q}_p
 \times \prod_{p\not\in {\mathcal P}} {\mathbb Z}_p \, .         \label{2.4}
\end{equation}

A multiplicative group of ideles ${\mathbb I} \subset {\mathbb A}$
has elements $x= (x_\infty, x_2, \cdots, x_p, \cdots)$,  where
$x_\infty \in {\mathbb R}^\ast = {\mathbb R} \setminus \{ 0\}$ and
$x_p \in {\mathbb Q}^\ast_p = {\mathbb Q}_p \setminus \{0 \}$ with
the restriction that for all but a finite set $\mathcal P$  one has
that $x_p \in {\mathbb U}_p = \{y \in \mathbb{Q}_p \, : |y|_p = 1
\}$. Thus the whole set of ideles is
\begin{equation}
 {\mathbb I} = \bigcup_{{\mathcal P}} {\mathbb I} ({\mathcal P}),
 \ \ \ \ {\mathbb I} ({\mathcal P}) = {\mathbb R}^{\ast}\times \prod_{p\in {\mathcal P}}
 {\mathbb Q}^\ast_p
 \times \prod_{p\not\in {\mathcal P}} {\mathbb U}_p \, .         \label{2.5}
\end{equation}

A principal adele (idele) is a sequence $ (x, x, \cdots, x, \cdots)
\in {\mathbb A}$ , where $x \in  {\mathbb Q}\quad (x \in {\mathbb
Q}^\ast = {\mathbb Q}\setminus \{ 0\})$. ${\mathbb Q}$ and ${\mathbb
Q}^\ast$ are naturally embedded in  ${\mathbb A}$ and ${\mathbb I}$
, respectively.

An additive character on $\mathbb{A}$ is
\begin{equation}
\chi (x) = \chi_\infty (x_\infty) \prod_p  \chi_p  (x_p) = \exp (- 2
\pi i x_\infty)  \prod_p  \exp (2 \pi i \{ x_p\}_p)\,, \quad  x \in
\mathbb{A} \,, \label{2.6}
\end{equation}
where $\{ x_p \}_p$ is the fractional part  of $x_p$ . A
multiplicative character on $\mathbb{I}$ is
\begin{equation}
|x|^s = |x_\infty|^s_\infty  \prod_p  |x_p|_p^s \,, \quad  s\in
\mathbb{C} \,, \quad x\in \mathbb{I}\,,    \label{2.7}
\end{equation}
where $\mathbb{C}$ is the field of complex numbers and
$|\cdot|_\infty$ denotes standard absolute value. One can easily see
that only finitely many factors in (\ref{2.6}) and  (\ref{2.7}) are
different from unity.

\section{Adelic quantum mechanics}         

Adelic quantum mechanics (see, e.g. \cite{dragovich1},
\cite{dragovich2}, \cite{dragovich3}) can be defined as a triple
$(L_2 (\mathbb{A})\,, W (z)\,, U (t))$, where $L_2 (\mathbb{A})$ is
the Hilbert space of complex-valued square integrable functions with
respect to the Haar measure on $ \mathbb{A}$,  $W (z)$ is a unitary
representation of the Heisenberg-Weyl group on $L_2 (\mathbb{A})$,
and $U (t)$ is a unitary representation of the evolution operator on
$L_2 (\mathbb{A})$.

A basis of $L_2 (\mathbb{A})$ can be the set of orthonormal
eigenfunctions in a spectral problem of the evolution operator $U
(t) \,, \, t \in \mathbb{A}$. Such eigenfunctions have the form
\begin{equation}
\psi_{\mathcal{P}} (x, t) = \psi_\infty (x_\infty , t_\infty)
\prod_{p \in \mathcal{P}} \psi_p (x_p , t_p) \prod_{p \not \in
\mathcal{P}} \Omega (|x_p|_p) \,, \quad x, t \in \mathbb{A} \,,
\label{2.8}
\end{equation}
where $\psi_\infty \in L_2 (\mathbb{R})$ and $\psi_p \in L_2
(\mathbb{Q}_p)$ are eigenfunctions in the ordinary and $p$-adic
cases, respectively. Eigenfunction $\Omega (|x_p|_p)$ is defined as
a characteristic function of $\mathbb{Z}_p$, i.e.
\begin{equation}
\Omega (|x_p|_p) = \left\{  \begin{array}{ll}
                 1,   &   |x_p|_p \leq 1,  \\
                 0,   &   |x_p|_p > 1 .
                 \end{array}    \right.
                 \label{2.9}
\end{equation}
This $\Omega (|x_p|_p)$ provides convergence of the infinite product
in (\ref{2.8}) and is invariant under transformation of $p$-adic
evolution operator $U (t_p)$.

In AQM quantization performs according to the Weyl procedure.

An adelic evolution operator is defined \cite{dragovich3} by
\begin{equation} \nonumber
U(t'')\, \psi(x'')=\int_{{\mathbb{A}}} {\cal K}(x'',t'';x',t')\,
\psi(x',t')\, dx' =\prod\limits_{v}{} \int_{{\mathbb Q}_{v}}{\cal
K}_v (x''_{v},t''_v;x'_v,t'_v)\, \psi_v(x'_v,t'_v)\, dx'_v,
\label{2.10}
\end{equation}
where $v=\infty, 2, 3,\cdots, p,\cdots$. The eigenvalue problem for
$U(t)$ reads
\begin{equation} U(t)\, \psi _{\mathcal{P}} (x)=\chi
(E\, t)\,  \psi _{\mathcal{P}} (x) \,, \quad x,t,E \in \mathbb{A}
\,, \label{2.11}
\end{equation}
where $\psi_{\mathcal{P}} (x)$ are adelic eigenfunctions
(\ref{2.8}). Since all information on quantum dynamics may be
derived from the kernels ${\cal K}_v (x''_{v},t''_v;x'_v,t'_v)$ they
can be regarded as basic ingredients of AQM. It is natural to
evaluate ${\cal K}_v (x''_{v},t''_v;x'_v,t'_v)$ by Feynman's path
integral, which for quadratic systems is defined by
\begin{equation}
{\cal K}_v (x''_{v},t''_v;x'_v,t'_v) = \int \chi_v \Big(
-\frac{1}{h} \int_{t'_v}^{t''_v} L (\dot{q} , q , t ) dt \Big) \,
\mathcal{D}_v q \,.    \label{2.12}
\end{equation}

 As a result of adelic approach and $p$-adic effects in AQM one
obtains discreteness of space at the characteristic (Planck) scale.
In the limit of large distances, AQM effectively becomes the
ordinary one. AQM may be regarded as a starting point for more
complete quantum cosmology, quantum field theory and
string/M-theory.

\subsection{Adelic form invariance}

In mathematical models of physical systems number sets which have
field structure play a very important role. In 1987 Volovich
\cite{volovich} suggested that a fundamental physical theory should
be formulated in such way that it is invariant under change of any
number field. So far it has not been constructed such number field
invariant quantum model. Especially it seems difficult to have a
theory simultaneously invariant on fields which are completions of
$\mathbb{Q}$ and finite Galois fields. However if we require
invariance only with respect to $\mathbb{R}$ and $\mathbb{Q}_p$ then
adelic quantum mechanics satisfies such requirement.

To illustrate this kind of invariance let us consider exact
expression for kernel (\ref{2.12}) for quantum-mechanical systems
with quadratic Lagrangian
\begin{equation}
L (\dot{q}\,, q\,, t) = \frac{1}{2}\, A(t)\,\dot{q}^2 + B(t)\,
\dot{q}\, q + \frac{1}{2}\, C(t)\, q^2 + D(t)\, \dot{q} + E(t)\, q +
F(t) \,, \label{2.13}
\end{equation}
where  $A (t), \cdots , F (t)$ are some analytic functions of the
time $t$ with rational coefficients. It is worth noting that
(\ref{2.13}) may be regarded as real as $p$-adic and it yields
quadratic classical action $\bar{S} (x'',t''; x',t')$. The
corresponding $v$-adic kernel \cite{dragovich5} is
\begin{equation}
{\cal K}_v (x'', t''; x', t') = \lambda_v \Big( -\frac{1}{2 h}
\frac{\partial^2}{\partial x'' \, \partial x'} \bar{S} \Big) \,
\Big| \frac{1}{h} \frac{\partial^2}{\partial x'' \, \partial x'}
\bar{S} \Big|_v^{\frac{1}{2}} \, \chi_v \Big( -\frac{1}{h}
\bar{S}(x'', t''; x', t') \Big) \,,    \label{2.14}
\end{equation}
where we omitted index $v$ in arguments because it is understood.
Arithmetic functions $\lambda_v$ are well defined complex-valued
functions (see, e.g. \cite{vladimirov1}).

It is now easy to observe that ${\cal K}_v (x'', t''; x', t')$ in
(\ref{2.14}) has the same form for any $v = \infty, 2, 3, \cdots, p
\cdots$, i.e. it is form invariant under interchange of any two
values of index $v$, which characterize number fields $\mathbb{R}
\equiv \mathbb{Q}_\infty$ and $\mathbb{Q}_p$.

Also the space of adeles $\mathbb{A}$ is  defined (\ref{2.4}) in
$v$-adic number field invariant way. Moreover, we see here that not
only fields of numbers are important but also rings $\mathbb{Z}_p$
of $p$-adic integers. Thus adelic space is also form invariant under
interchange of any two rings $\mathbb{Z}_{p'}$ and
$\mathbb{Z}_{p''}$ as well as under interchange of $\mathbb{Q}_{p'}$
and $\mathbb{Z}_{p''}$. It follows that in adelic theory
$\mathbb{Z}_{p}$ is not less important than $\mathbb{Q}_{p}$.

Eigenfunctions (\ref{2.8}) make also an  example of adelic form
invariance.  $\Omega (|x|_p)$ functions are a consequence of
$\mathbb{Z}_{p}$ structure of $\mathbb{A}$ and they are necessary
ingredient of AQM.

\subsection{Adelic rational invariance}      

Now we are interested in existence of  adelic quantities which are
invariant under some adelic transformations.  If we restrict
transformations to principal adeles or principal ideles then there
exist such invariant quantities. Since principal adeles and ideles
are related to rational numbers we shall call this kind of
invariance {\it adelic rational invariance} (ARI). The simplest case
of ARI gives the following example of adelic multiplicative
character (\ref{2.7})
\begin{equation}
|x| = |x|_\infty  \prod_p  |x|_p  = 1 \,, \quad x\in
\mathbb{Q}^\ast\,, \label{2.15}
\end{equation}
which is valued for any rational $x \neq 0$. One can easily show
validity of (\ref{2.15}) as well as its extension to $|x|^s =1$ if
$s \in \mathbb{C}$ and $x\in \mathbb{Q}^\ast$.

The next example of ARI is related to additive character
(\ref{2.6}). Namely
\begin{equation}
\chi (x) = \chi_\infty (x) \prod_p  \chi_p  (x) = \exp (- 2 \pi i x)
\prod_p  \exp (2 \pi i \{ x\}_p) = 1 \,, \quad  x \in \mathbb{Q} \,.
\label{2.16}
\end{equation}

One can show that adelic kernel
\begin{equation}
{\cal K} (x'', t''; x', t') = \prod_v  {\cal K}_v (x'', t''; x', t')
=1  \,, \quad  x'', x' , t'' , t' \in \mathbb{Q}^\ast
\,,\label{2.17}
\end{equation}
where ${\cal K}_v (x'', t''; x', t')$ is given in (\ref{2.14}), if
classical action $\bar{S} (x'', t''; x', t')$ is rational function
of its arguments. Proof that $\prod_v \lambda_v (x) =1$ if $x \in
\mathbb{Q}^\ast$ can be found in \cite{vladimirov1}.

It is worth noting that infinite products in (\ref{2.15}),
(\ref{2.16}) and (\ref{2.17}) are equal to unity. They connect real
and $p$-adic counterparts of the same quantity. As a result, real
quantity of rational arguments can be expressed as product of all
inverse $p$-adic analogues.

\section{Conclusion}

We presented some results on two kinds of  symmetries in adelic
quantum mechanics: form invariance and rational invariance. In
addition to number fields it is pointed out importance of the rings
of $p$-adic integers. These adelic quantum symmetries should
stimulate further developments in adelic approach to QFT and
string/M-theory.

\bigskip
{\small Ackowledgement.  The work on this article was partially
supported by the Ministry of Science and Environmental Protection,
Serbia, under contract No 144032D.}

\end {document}